 \documentclass[preprint,3p,times,twocolumn,authoryear]{elsarticle}

\usepackage{amssymb}
\usepackage{lipsum}
\usepackage{gensymb}

\usepackage{pdfpages}
\usepackage{caption}
\usepackage{url}
\usepackage{array}

\usepackage{setspace}
\usepackage{lineno}

\journal{Astronomy $\&$ Computing}

\usepackage{float}

\begin{document}
\makeatletter
\def\ps@pprintTitle{%
  \let\@oddhead\@empty
  \let\@evenhead\@empty
  \def\@oddfoot{\footnotesize\itshape Preprint submitted to Astronomy \& Computing\hfill\today}%
  \let\@evenfoot\@oddfoot}
\makeatother
\begin{frontmatter}



\title{Understanding the regulation of star formation within TNG100 galaxies on kpc-scales using machine learning I: Global versus local}

\author[AU,NEU]{Bryanne McDonough \corref{cor1}}
\cortext[cor1]{Corresponding Author}
\ead{bmcdonough@adelphi.edu}

\affiliation[AU]{organization={Physics Department, Adelphi University},
                addressline={1 South Avenue},
                city={Garden City},
                postcode={11530},
                state={NY},
                country={USA}}

\affiliation[NEU]{organization={Department of Physics, Northeastern University},
            addressline={360 Huntington Ave}, 
            city={Boston},
            postcode={02115}, 
            state={MA},
            country={USA}}

\author[NEU]{Sathvika S. Iyengar}

\author[NEU]{Ansa Brew-Smith}

\author[FIU]{Asa F. L. Bluck}

\affiliation[FIU]{ organization= {Stocker AstroScience Center, Dept. of Physics, Florida International University},
    addressline={11200 SW 8th Street},
    city={Miami},
    state={FL},
    postcode={33199},
    country={USA}
}

\author[CIT]{Joanna M. Piotrowska}
\affiliation[CIT]{organization={Department of Astronomy, California Institute of Technology},
    addressline={1200 East California Boulevard}, 
    city={Pasadena},
    state={CA},
    postcode={91125}, 
    country={USA}
}

\begin{abstract}
We apply Random Forest and \texttt{XGBoost} machine learning algorithms to determine which galaxy properties most effectively predict star formation and quenching in simulated galaxies. Using spatially-resolved data from approximately 63,000 annular bins across $6,189$ TNG100 galaxies, we train classification models to predict quenching states and regression models to predict star formation rate surface densities. Despite their different algorithmic approaches, both methods produce consistent feature importance rankings, with XGBoost distributing importance more evenly among correlated features. For central galaxies 
and high-mass satellites, 
black hole mass dominates quenching predictions, consistent with quenching via active galactic nuclei (AGN) feedback. Classification of low-mass satellites shows overwhelming importance for halo mass, indicating environmental quenching. 
Star formation predictions are dominated by local stellar mass surface density across all star-forming galaxy types, confirming that active star formation is a local process while quenching is driven by global properties. 
\end{abstract}




\end{frontmatter}






\section{Introduction}
\label{sec:introduction}

There exists a bimodality in the star formation properties of galaxies at low-redshift, with most galaxies either forming stars at a rate expected for their mass or lacking significant levels of star formation \citep[e.g.][]{2001AJ....122.1861S,Brinchmann04,2006MNRAS.373..469B,Peng10}. Galaxies in the latter category are often referred to as quenched or quiescent galaxies. The quenched fraction has been shown to depend on various intrinsic and environmental properties associated with galaxies, especially mass, environment, and morphology \citep[e.g.,][]{2006MNRAS.373..469B,2006MNRAS.368..414D,Martig,2009ApJ...699..105C,Peng10,Peng12,2014MNRAS.441..599B,2014MNRAS.440..843O,2014ApJ...788...11L}. 

Galaxies will continue to form stars as long as they have a supply of cool, dense molecular gas that can collapse into protostellar objects. Molecular gas in the interstellar medium (ISM) is constantly depleted as it is formed into stars. In high-mass star-forming galaxies, the ISM can be continually replenished via accretion from the circumgalactic medium (CGM). The ISM of low-mass galaxies can be replenished via streams of cold gas accreted from the intergalactic medium \citep[e.g.,][]{2005Natur.433..604D,2005MNRAS.361..776S,2006MNRAS.368....2D,2006ApJS..163....1H,2008ApJS..175..356H,2012MNRAS.425L..66M}.
There are several avenues through which galaxies may lose access to their reservoirs of star formation fuel. Winds from supernovae and/or active galactic nuclei (AGN) can eject gas from the ISM and into the CGM \cite[e.g.,][]{somerville2015ARA&A..53...51S,Naab2017ARA&A..55...59N, veilleux2020A&ARv..28....2V}. Accretion of fresh gas onto the ISM from the CGM can be prevented via heating of the circumgalactic gas or removal of the CGM via stripping and/or tidal interactions \citep[e.g.,][]{2006MNRAS.366..417F,2006MNRAS.365...11C,2007MNRAS.380..877S,Weinberger17,tumlinson2017ARA&A..55..389T,simpson2018MNRAS.478..548S}. More recent studies have also connected turbulence in the ISM to quenching \citep[e.g.,][]{Weinberger18,2020MNRAS.499..768Z,Piotrowska2020MNRAS.492L...6P,Piotrowska2022MNRAS.512.1052P}. Alternative quenching pathways proposed in the literature include halo mass quenching through virial shock heating of infalling gas \citep[e.g.,][]{2003MNRAS.345..349B,2016MNRAS.457.4360Z}, supernovae feedback \citep[e.g.,][]{2010MNRAS.403L..16C,2021ApJ...909..192G}, morpholgoical quenching via dynamical stabilization \citep[e.g.,][]{Martig,2023ApJ...953...27L}, and environmental quenching \citep[e.g.,][]{Peng12,2022A&ARv..30....3B,2014ApJ...789..164T,2016MNRAS.463.1916F,2017MNRAS.469.3670S,2022A&ARv..30....3B,McDonough2025ApJ...978...67M,2025MNRAS.543.2006G}.

Whether, and to what extent, these processes operate on galaxies and ultimately drive the cessation of star formation is still under debate. Several studies have indicated different quenching pathways for different galaxy populations. In particular, different pathways to quenching have been identified for central galaxies and low-mass satellite galaxies \citep[e.g.,][]{Peng12,Bluck2020b,2024MNRAS.528.4891G,2025MNRAS.543.2006G}. Satellite galaxies are more susceptible to environmental processes—such as galaxy-galaxy tidal interactions and ram pressure stripping—as they orbit within the halos of more massive centrals. 

While it has become clear that a variety of galaxy parameters are correlated with quenching, the relative importance of these parameters in the quenching process is still unclear. For example, the higher quenched fractions of satellite galaxies in massive clusters indicates that the quenching of these systems is likely driven by environmental interactions. Potential drivers of environmental quenching include tidal stripping from galaxy-galaxy interactions \citep[e.g.,][]{1962AJ.....67..471K,Wilkinson21} and ram pressure stripping from galaxy-halo interactions \citep[e.g.,][]{1972ApJ...176....1G,2007A&A...472....5J,2009A&A...500..693J,2022A&ARv..30....3B}. Which processes dominate the quenching of low-mass satellite galaxies in massive clusters remains an open question. 

The primary challenge in understanding which features of a galaxy drive quenching is that many of these features are intercorrellated. For example, the mass of the supermassive black hole at the center of a galaxy is correlated with the stellar mass of that galaxy, which is in turn correlated with the mass of the gaseous halo surrounding the galaxy. This complication has been addressed in prior works \citep[e.g.,][]{Bluck2020a,Bluck2020b,Piotrowska2020MNRAS.492L...6P,Piotrowska2022MNRAS.512.1052P,2024MNRAS.528.4891G,2025MNRAS.543.2006G} with a machine learning tool known as Random Forest \citep[RF; ][]{Breiman2001,RF10.1007/978-3-642-34062-8_32,scikit-learn}. Random Forest classification utilizes an ensemble of decision trees to determine the relative importance of various `features' (i.e., properties) of galaxies for predicting whether a galaxy is quenched or star forming. Using a Random Forest classifier on spatially-resolved observational data, \cite{Bluck2020a} found that quenching of central galaxies was best predicted by global (i.e., integrated) features. In a similar analysis, \cite{Bluck2020b} found that quenching of regions from observed central galaxies and high-mass satellite galaxies was best predicted by the mass of the central supermassive black holes (as inferred by central velocity dispersion) while the influence of environmental parameters, particularly the local galaxy overdensity, was much more pronounced for the quenching of low-mass satellites. \cite{2024MNRAS.528.4891G} found nearly identical results in an analysis of integrated properties of SDSS galaxies. However, they found that halo mass, rather than overdensity, was a better predictor of quenching for simulated low-mass satellite galaxies. 

Accurately modeling the quenching of galaxies in hydrodynamic simulations of galaxy evolution has historically been a challenge. However, with careful implementation of feedback models, modern simulations are now reproducing observationally-realistic galaxy populations. Of concern is how well these feedback models actually capture real physical processes. Comparing the relative importance of various features for quenching in both observed and simulated galaxies offers another way to analyze the accuracy of feedback models. While \cite{2024MNRAS.528.4891G} compared Random Forest analyses of observed galaxies from SDSS and simulated galaxies from Illustris \citep{Nelson15,Vogelsberger14} , TNG100 \citep{Nelson2018,2018MNRAS.475..648P,2018MNRAS.475..676S,2018MNRAS.477.1206N,2018MNRAS.480.5113M,Nelson2019a}, and EAGLE \citep{Crain15, 2015MNRAS.446..521S}, this work looked only at properties on a global (i.e., integrated) level. However, quenching does not happen simultaneously at all points within a galaxy, instead some regions can be quenched while others are still actively star-forming. 
Thus, it is of interest to analyze the importance of various features at a spatially-resolved scale with machine learning methods that allow us to effectively control for spurious inter-correlations. \cite{Bluck2020a} and \cite{Bluck2020b} performed such an analysis on observed galaxies from MaNGA, but a similar analysis has not been performed on spatially-resolved data from simulations.

In this work, we perform a machine learning analysis on data from spatially-resolved galaxies in the TNG100 simulation. In addition to RF classification and regression, we explore a gradient-boosting machine learning technique, \texttt{XGBoost} \citep{scikit-learn}. 
Our results will be compared to results from a similar analysis of observed galaxies \citep{Bluck2020b}, which will be a further test of how well TNG100 feedback models regulate star formation and quenching on kpc-scales. Of particular interest is whether the tension between observations and simulations regarding which environmental parameter best predicts quenching of low-mass satellite galaxies found by \cite{2024MNRAS.528.4891G} is still present when the TNG100 galaxies are examined on spatially resolved scales.

The code used for this project has been made public on Github\footnote{\url{https://github.com/bryannemcd/tng_quenching_ml}} and the specific version of the code used for this paper is archived with Zenodo  [dataset]\citep{iyengar_2026_19236037}. The repository includes binned data, but the original spaxel-level data is also archived on Zenodo \citep{letterdata}.

\section{Data} 
\label{sec:data}
Our sample of galaxies is identical to that in \cite{McDonough2023ApJ...958...19M} and \cite{McDonough2025ApJ...978...67M}. These galaxies come from the $z=0$ snapshot of the highest resolution realization of the TNG100 simulation. TNG100 is a cosmological, magnetohydrodynamic simulation with a comoving box length of $75 h^{-1}\textrm{Mpc} $ and cosmology from \cite{Planck2016}. The simulation has a target minimum baryonic mass of $\sim 10^6 h^{-1}M_\odot$. 

The selected galaxies have at least $1000$ stellar particles, resulting in a minimum mass of $\approx 10^9 M_\odot$. For this analysis, we divide our sample by mass and categorization into central or satellite galaxy. Central galaxies are defined to be the most massive galaxies in their friends-of-friends (FoF) halos, while less massive members of the FoF halo are considered satellites. 
We separate high- and low-mass galaxies at a stellar mass of $2\times10^{10} M_\odot$, the mass-quenching threshold identified for TNG galaxies by \cite{2025MNRAS.543.2006G}.
Our high-mass samples include $2520$ central galaxies and $969$ satellite galaxies. Our low-mass samples include $2085$ central galaxies and $624$ satellite galaxies.

From the simulation group catalogs, we adopt the total stellar mass of all bound particles as $M_*$, the central black hole mass $M_{\rm BH}$, and the total mass of all particles bound to the friends-of-friends halo as the halo mass $M_H$. 
We compute the local overdensity at the fifth nearest neighbor, $\delta_5$, as:
\begin{equation}
    \delta_5 \equiv \frac{n_5}{\bar{n}}-1 = \frac{5}{\frac{4}{3} \pi d_5^3} \frac{L_{\mathrm{box}}^3}{N_{\mathrm{total}}}-1,
\end{equation}
where $d_5$ is the distance of a galaxy from its fifth nearest neighbor, $L_{\rm box}$ is the simulation box length, and $N_{\rm total}$ is the total number of galaxies in the box with an absolute r-band magnitude $M_r<-20$.\footnote{This magnitude cut is selected to be consistent with the procedure for calculating nearest neighbor overdensities in SDSS data presented in \cite{2006MNRAS.373..469B}. } 

For each galaxy, we map bound particles within a distance of $2$ times the half-light radius\footnote{We adopt half-light radii for TNG galaxies from the Stellar Projected Sizes catalog presented in \cite{2018MNRAS.474.3976G}.} to a grid using the standard cubic-spline kernel, with a grid spacing of $0.5 h^{-1} {\rm kpc}$. These grid bins are meant to mimic observational `spaxels' or spectral pixels, so we will refer to them as such. The mapping procedure is described in detail in \cite{McDonough2023ApJ...958...19M}. We generate maps of stellar mass density, $\Sigma_*$ and time-averaged star formation rate density, $\Sigma_{\rm SFR}$.
We limit our analysis to spaxels where $\Sigma_*>10^6 M_\odot {\rm kpc}^{-2}$. Star formation rates are calculated from stars formed over the last $20$ Myr or, where no stars have formed in that time, over the last $100$ Myr. 
These timescales are meant to approximate the timescales of the observational H$\alpha$ and $4000 \AA$ break SFR tracers, respectively. 
Due to the limited baryonic mass resolution of the simulation, there are often spaxels where no stars have formed in the last $100$ years. 
In this case, we assume a specific SFR of ${\rm sSFR} = 10^{-12} \; {\rm yr}^{-1}$, with scatter drawn from a normal distribution with width $\sigma=0.1$ dex. 

\subsection{Pre-processing}


The resulting maps are often patchy, which would confuse the machine learning models we describe below. Therefore, we bin the spaxels into annuli with a width of $1.5 \; {\rm kpc}$, which is approximately twice the width of individual spaxels. Altering the bin size does not significantly impact our results. Given that the outermost bin of each galaxy may be incompletely filled, we exclude each of those bins from our analysis.


To classify the annular bins as quenched or star-forming, we compare them to a resolved star-forming main sequence (rSFMS) fit obtained from \texttt{ScaleRPy} \citep{2025ApJ...986L..32M}. To avoid fitting to quenched regions, we re-bin using a subset of our spaxels which excludes spaxels with SFRs imposed via the lower limit. That is, the bins used to identify the rSFMS do not include quenched spaxels. This does not significantly impact the resulting rSFMS slope because we do not consider the area of the removed spaxels when computing the SFR density. We fit the rSFMS to a single linear function (in base-10 log space) for consistency with previous works. The fit results in an rSFMS that is described by:
\begin{equation}
    \centering
    \log_{10} \Sigma_{\mathrm{SFR,MS}} = (0.79 \pm 0.02)\times \log_{10} \Sigma_* - (7.6 \pm 0.1).
\end{equation}


Our bins that use the full set of spaxels can then be classified as quenched or star forming
based on their logarithmic offset from the rSFMS, $\Delta \Sigma_{\mathrm{SFR}}$:
\begin{equation}
    \Delta \Sigma_\mathrm{SFR}=\log_{10}(\Sigma_{\mathrm{SFR}}) - \log_{10}(\Sigma_{\mathrm{SFR,MS}}).
\end{equation}
This offset considers both the stellar mass surface density and star formation rate surface density of each bin.
A bin is classified as quenched when $\Delta \Sigma_{\rm SFR} < -2.4$, otherwise it is considered star-forming, following \cite{Bluck2020b}.
In total, we had $21,983$ quenched bins and $41,099$ star-forming bins. 

Each annular bin inherits global properties (e.g., central black hole mass, local overdensity) from its parent galaxy. Each bin is also associated with its radial distance from the center of the galaxy and its local stellar mass surface density. Including both global and local variables helps avoid having a dataset with many duplicate or contradictory entries.
We note there are some TNG100 galaxies in our sample which do not host a supermassive black hole. In these cases we assign the galaxy a black hole mass of $0$. In the simulation, a black hole is seeded in the center of a galaxy if 1) its host halo exceeds a mass threshold and 2) the halo does not already host a black hole. In our data set, there are only $8$ central galaxies with $M_{\rm BH}=0$, all low-mass galaxies that have not exceed the halo mass threshold. There are $99$ satellites with $M_{\rm BH}=0$, only $5$ of which are high-mass. In these cases, the satellites must not have exceeded the halo mass threshold prior to joining their host halo. Once in their host halo, they would not have been seeded with a black hole because the halo they joined already hosted one. 



The bins were then sorted into four subsets which were analyzed by the machine learning models: low-mass satellites, low-mass centrals, high-mass satellites, high-mass centrals. 


\begin{table*}[ht!]
\centering
\begin{tabular}{|p{1.5cm}|p{1.5cm}|p{1.5cm}|p{1.5cm}|p{1.5cm}|}
\hline
 & \multicolumn{2}{c|}{\textbf{Centrals}} & \multicolumn{2}{c|}{\textbf{Satellites}} \\
\cline{2-5}
 & {Low-Mass} & {High-Mass} & {Low-Mass} & {High-Mass} \\
\hline
\textbf{Quenched} & $3,370$ & $12,253$ & $1,867$ & $4,493$ \\
\hline
\textbf{Star-forming} & $18,307$ & $14,646$ & $3,964$ & $4,182$ \\
\hline
\end{tabular}
\caption{Number of quenched and star-forming bins for all four subgroups}
\end{table*}

\begin{table*}[ht!]
\centering
\begin{tabular}{|l|l|l|}
\hline
\textbf{Hyperparameter} & \textbf{Range / Value} & \textbf{Definition \citep{scikit-learn}}\\
\hline
\texttt{n\_estimators} & [100--700]  & \texttt{Number of trees}\\
\hline

\texttt{max\_depth} & [10--250] & \texttt{Maximum tree depth}\\
\hline

\texttt{max\_samples\_split} & [50--400] & \texttt{Minimum number of samples to split internal node}\\
\hline

\texttt{min\_samples\_leaf} & [100--700] & \texttt{Minimum number of samples to split leaf node}\\
\hline

\texttt{max\_features} & \texttt{"None"} & \texttt{Maximum number of features at split}\\
\hline

\texttt{bootstrap} & \texttt{"True"} & \texttt{Use bootstrap samples for building trees}\\
\hline

\texttt{class\_weight} & \texttt{"balanced"} & \texttt{Weights for each class}\\
\hline
\end{tabular}
\caption{Hyperparameter values used in RF model} \label{tab:RF}
\end{table*}

\begin{table*}[ht!]
\centering
\begin{tabular}{|l|l|l|}
\hline
\textbf{Hyperparameter} & \textbf{Range / Value}  & \textbf{Definition \citep{scikit-learn}}\\
\hline
\texttt{n\_estimators} & [100--800] & \texttt{Number of gradient boosted trees}\\
\hline

\texttt{max\_depth} & [1--3] & \texttt{Maximum tree depth}\\
\hline

\texttt{learning\_rate} & [0.003--0.05] & \texttt{Boosting learning rate, $\eta$} \\
\hline

\texttt{subsample} & [0.5--0.8] & \texttt{Subsample ratio of training instance}\\
\hline

\texttt{colsample\_bytree} & 1.0 & \texttt{Subsample ratio of feature columns for each tree}\\
\hline

\texttt{min\_child\_weight} & [20--120] & \texttt{Minimum sum of instance weight needed in a child node}\\
\hline

\texttt{gamma} & [0.2--1.5] & \texttt{Minimum loss reduction need for leaf node partition}\\
\hline

\texttt{alpha} & [4--15] & \texttt{L1 regularization term}\\
\hline

\texttt{lambda} & [4-25] & \texttt{L2 regularization term}\\
\hline

\texttt{scale\_pos\_weight} & 1.0 & \texttt{Balance positive and negative weights}\\
\hline
\end{tabular}
\caption{Hyperparameter values used in XG model} \label{tab:XGB}
\end{table*}

\section{Methods}
\label{sec:methods}
\begin{sloppypar}
Following the methods detailed in \cite{Bluck2020a,Bluck2020b}, we develop Random Forest (RF) models to analyze the importance of various features for predicting whether regions of TNG100 galaxies are quenched or star-forming. We compare this method to an alternative algorithm, \texttt{XGBoost} \cite[]{Chen:2016:XST:2939672.2939785}. The main difference between these models is that \texttt{XGBoost} uses a Boosting technique while RF uses a Bagging technique. That is, RF builds decision trees independently and in parallel, while \texttt{XGBoost} builds decision trees sequentially to iteratively lower error.
We draw our classification and regression tools from the \texttt{scikit-learn} package \cite[]{scikit-learn} for \texttt{Python}.
\end{sloppypar}

RF and XGBoost are ensemble methods, meaning they combine multiple models to improve the prediction accuracy. Both methods provide feature importance, indicating how much each feature provided in the data contributes to the overall decision.

\subsection{Training, Testing, and Tuning} \label{sec:tuning}

For classification models, we evaluate performance using the Area Under the Receiver Operating Characteristic Curve (AUC-ROC), which measures the model’s ability to distinguish between classes by plotting the true positive rate (TPR) against the false positive rate (FPR) \citep[e.g.,][]{10.1093/mnras/stw036, 10.1093/mnras/stz363, Bluck2020a}. For regression models, we use Mean Squared Error (MSE), defined as the average squared difference between predicted and measured $\Sigma_{\rm SFR}$ values \citep[following ]{Bluck2020a}. We compute the MSEs on raw values, without any scaling.

To verify the efficacy of each model on unseen data, the datasets are split into training and testing samples, each containing $50\%$ of the galaxies in a given sample. We prevent data leakage by splitting on galaxy IDs, so that bins belonging to the same galaxy will be in either training or testing sets, never both in the same run. 

For both RF and \texttt{XGBoost} analyses of each galaxy category, we conduct $10$ independent runs, which are randomized with different \texttt{random\_state} seeds. The random seed is the source of variation for each run, as it affects the train/test split, model, and hyperparameter cross validation. Thus, each independent run uses different sets of galaxies for training and testing. 

In typical use cases for classification and regression models, it is desirable to achieve the best possible AUC or MSE scores on the tested (unseen) dataset. High (low) AUC (MSE) scores indicate that the models are able to predict the desired characteristic with high accuracy. Our use case differs in that we already know the desired characteristic ($\Sigma_{\rm SFR}$ or classification as quenched or star forming), but wish to know which parameters are most important for making that prediction. To optimize the model for this case, we perform coarse hyperparameter tuning by manually adjusting the parameter grid used by the models for each galaxy subsample in order to achieve a maximum difference of $0.03$ between training and testing AUC/MSE scores over the $10$ independent runs whilst maximizing the permanence within this constraint. Within each independent run, RF and \texttt{XGBoost} use \texttt{RandomSearchCV} to fine tune the hyperparameters on the training set.

We take as our uncertainty the variance of importance scores assigned between independent runs.

\subsection{Random Forest}

\begin{sloppypar}

RF utilizes the Bagging (bootstrap aggregation) technique, which builds decision trees on different subsets from the same dataset, taking the majority prediction for classification and average for regression \cite[]{KHAN2024122778}. Decision trees follow a tree-like structure, where decisions are made to minimize node impurity using metrics such as entropy or Gini impurity. For more details regarding how a Random Forest algorithm works and applications to astronomical data, we direct readers to Appendix B of \cite{2022A&A...659A.160B}.

From \texttt{scikit-learn} \citep{scikit-learn}, we use \texttt{RandomForestClassifier} for classification and \texttt{RandomForestRegressor} for regression. In Table \ref{tab:RF}, we provide the range of hyperparameter values adopted after the coarse tuning described in \S \ref{sec:tuning}. For the precise hyperparameter ranges used for each galaxy subset, we direct readers to the archive of our code linked above.
\end{sloppypar}

\begin{sloppypar}
Following \cite{Bluck2020b}, we vary \texttt{min\_samples\_leaf}, the minimum number of samples required at the leaf node, which mitigates overfitting. By default, RF does not evaluate every parameter at every split of each decision tree. As discussed in \cite{2022A&A...659A.160B}, this is undesirable if there is  inter-correlation between features. Thus, we use the 'All Parameter' mode of RF by setting \texttt{max\_features = None}.
\end{sloppypar}

The feature importance scores we report from RF models are calculated using the mean decrease in Gini impurity.

\subsection{XGBoost}
\texttt{XGBoost} is a type of Gradient Boosting method that adds weak prediction models in order to correct the errors made by previous models, to develop a stronger prediction model. For classifiers, the loss function to be minimized is typically binary cross-entropy, whereas for regressors, the loss function is typically mean-squared error \citep{KHAN2024122778}. Regularization techniques, such L1 (Lasso) and L2 (Ridge) regularization, are used to prevent overfitting .

\begin{sloppypar}
From \texttt{scikit-learn} \citep{scikit-learn}, we use \texttt{XGBClassifier} for classification and \texttt{XGBRegressor} for regression. 
For hyperparameter tuning, \texttt{RandomizedSearchCV} was used for classification and \texttt{GridSearchCV} was used for regression. Unlike \texttt{RandomizedSearchCV}, \texttt{GridSearchCV} performs a more exhaustive search, iterating through every possible combination for optimal performance. In Table \ref{tab:XGB}, we report the values of hyperparameters used to obtain the results presented below.
\end{sloppypar}

\begin{sloppypar} 
Critically, we ensure all the features are available to each tree in \texttt{XGBoost} by setting \texttt{colsample\_bytree=1.0}. This is intended to mimic \texttt{max\_features= None} in RF, but is limited in that it only makes all features available to the model, rather than ensuring that the algorithm evaluates each feature at each split. In addition, L1 and L2 regularization terms were used as a means of reducing overfitting. 
\end{sloppypar}

The feature importance scores we report from \texttt{XGBoost} models are calculated via the default `weight,' the number of times a feature is used to split the data across all trees. We note that gain-based importance may be more comparable to importance scores from RF, but we did not observe a significant difference in our results when the gain-based importance was adopted.

\section{Results}
\label{sec:results}

Here, we present the mean feature importance scores averaged over 10 independent runs in both classification and regression algorithms. We will report mean feature importance scores for feature $X$ as $I_{\rm RF}(X)$ or $I_{\rm XGB}(X)$ for scores obtained from the Random Forest and \texttt{XGBoost} models, respectively. The reported uncertainties on these scores represent the standard deviation of the results from independent runs.

We classify features by whether they are local, intrinsic global, or environmental global. Local features are those unique to a given bin, and include the radial location of the annular bins normalized by effective radius ($r/R_e$) and the local stellar mass surface density ($\Sigma_{{*}}$). `Global' features are those that are unique to the galaxy that a bin belongs to. Intrinsic features are those that come from internal properties of a galaxy, while environmental features relate to the environment a galaxy resides in. Intrinsic global features include the global stellar mass  ($M_{{*}}$) and black hole mass ($M_{{BH}}$). 
The remaining features, halo mass ($M_{{h}}$) and local galaxy overdensity measured at the fifth nearest neighbor ($\delta_{{5}}$), make up the environmental global features. 

We note that halo mass is not precisely an environmental parameter for central galaxies, due to the tight correlation between halo mass and galaxy properties. However, for consistency we describe halo mass as an environmental parameter for each category of galaxies.


\subsection{Classification}

\begin{figure*}[ht]
	\centering 
	\includegraphics[width=\linewidth]{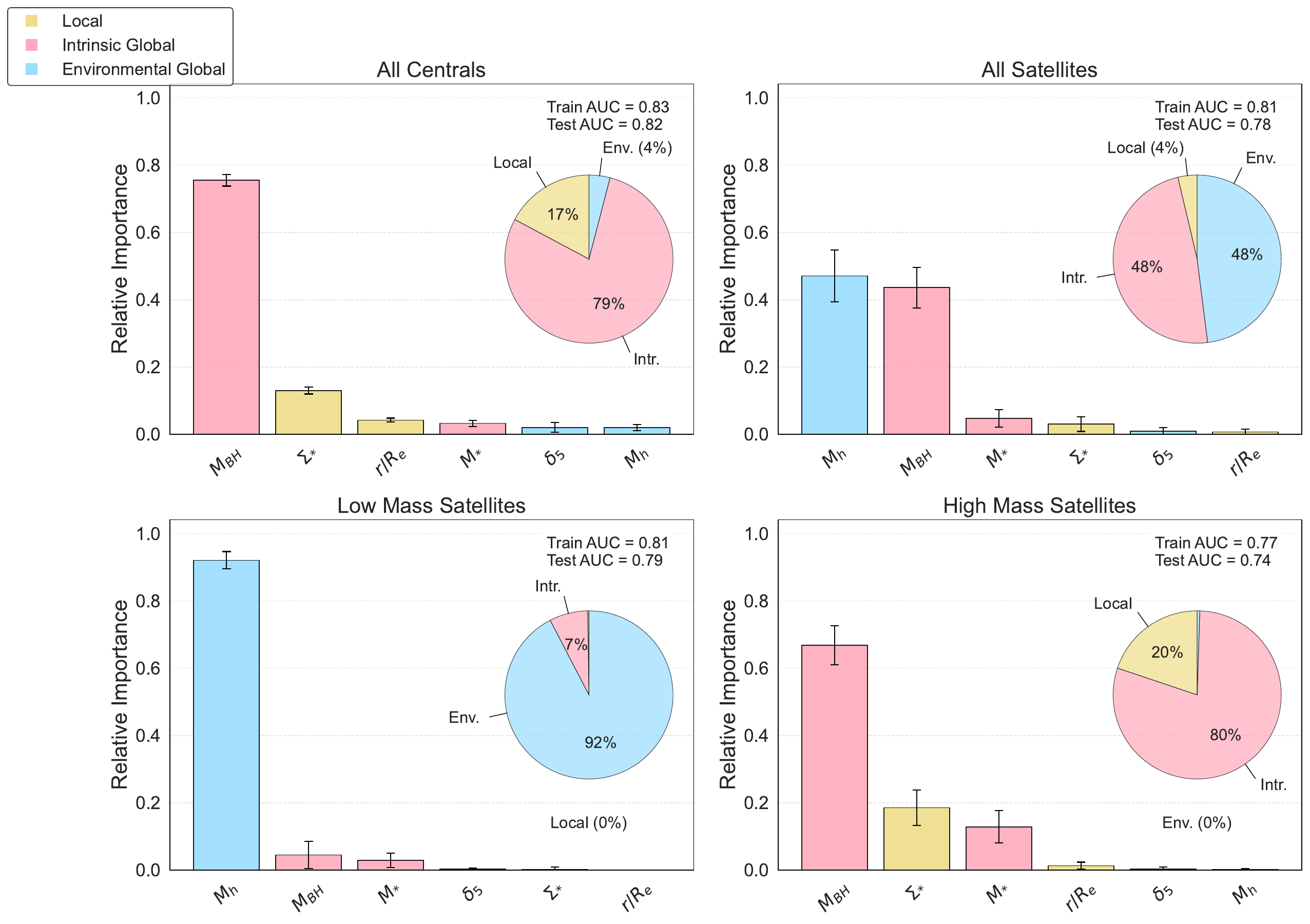}
	\caption{Average feature importances from the Random Forest quenching classification models, shown for spatially-resolved bins drawn from all centrals (top left), all satellites (top right), low-mass satellites (bottom left), and high-mass satellites (bottom right). In each panel, features are ranked from most to least important. The error bars represent variance over ten independent runs. Features that are local (i.e., unique to a bin) are colored yellow. Features that are global (i.e., unique to a galaxy) are colored pink or blue depending on whether they are an intrinsic feature or environmental feature, respectively. The inset pie charts sum the importance scores in each group to show the percent contributions of local, intrinsic global, environmental global. Each panel also includes the training and testing AUC scores averaged over ten independent runs. } 
	\label{fig:RF_classification}%
\end{figure*}

\begin{figure*}[ht]
	\centering 
	\includegraphics[width=\linewidth]{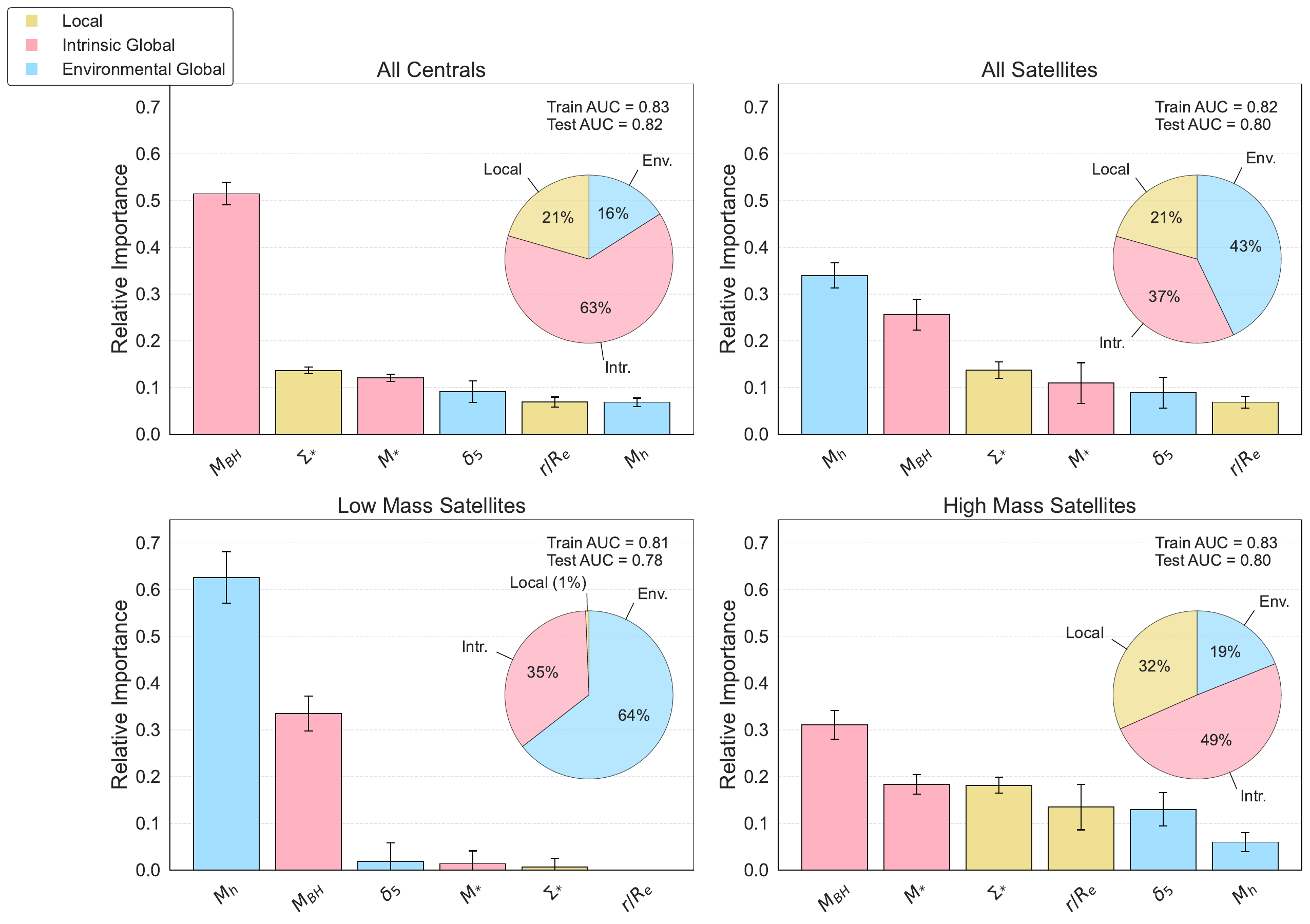}
	\caption{Feature importances from XGBoost quenching classification models, shown for low-mass and high-mass subsets of both centrals and satellites. Formatting is identical to Figure \ref{fig:RF_classification}.} 
	\label{fig:XGB_classification}%
\end{figure*}

Figures \ref{fig:RF_classification} and \ref{fig:XGB_classification} summarize results from the Random Forest and \texttt{XGBoost} classification algorithms, respectively. For ease of interpretation, local features are in yellow, global intrinsic features are in pink, and global environmental features are in blue. The inset pie charts summarize the total feature importance contributions for each category. 

Both algorithms are in agreement regarding the most important feature for each category of galaxy. However, the importance score of the most important feature varies between algorithms, with \texttt{XGBoost} generally returning lower scores. This is due to \texttt{XGBoost} finding higher scores for less important parameters as compared to the RF results. In \S \ref{sec:discussion}, we discuss how this is likely a result of how well the algorithms handle highly correlated parameters.

\begin{sloppypar}
    
Both algorithms are in agreement that intrinsic features are most predictive of quenching for all galaxy categories except low-mass satellites. Unsurprisingly, the $M_{\rm BH}$ feature dominates for central galaxies, with importance scores of $I_{\rm RF}(M_{\rm BH})=75\pm2\%$ and $I_{\rm XGB}(M_{\rm BH})=51\pm2\%$. The Random Forest models find that the only other parameter with importance $>10\%$ for central galaxies is $\Sigma_*$, with $I_{\rm RF}(\Sigma_*)=13\pm2\%$. The \texttt{XGBoost} models also find $\Sigma_*$ to be second-most important for centrals ($I_{\rm XGB}(\Sigma_*)=13.6\pm 0.7 \%$), but it is closely followed by $M_*$ with $I_{\rm XGB}(M_*)=12.0\pm 0.8 \%$.

For all satellite galaxies, both algorithms find that the top two parameters are $M_h$ and $M_{\rm BH}$ ($I_{\rm RF}(M_h)=47\pm8\%$; $I_{\rm XGB}(M_h)=34\pm3\%$; $I_{\rm RF}(M_{\rm BH})=44\pm6\%$; $I_{\rm XGB}(M_{\rm BH})=26\pm3\%$). This is unsurprising given that the bottom panels of Figures \ref{fig:RF_classification} and \ref{fig:XGB_classification} show that $M_h$ is the most important parameter for low-mass satellites and $M_{\rm BH}$ is the most important parameter for high-mass satellites. 

The bottom left panels of Figures \ref{fig:RF_classification} and \ref{fig:XGB_classification} show that $M_h$ is the dominate factor driving quenching in low-mass satellite galaxies ($I_{\rm RF}(M_h)=92\pm3\%$; $I_{\rm XGB}(M_h)=63\pm6\%$). While $M_{\rm BH}$ has little importance in the Random Forest models ($I_{\rm RF}(M_{\rm BH})=5\pm4\%$), \texttt{XGBoost} found a much higher importance for black hole mass ($I_{\rm XGB}(M_{\rm BH})=34\pm4\%$). This difference likely arises due to the different algorithms underlying Random Forest and \texttt{XGBoost}, which we will discuss further below. Of all galaxy categories we explore here, low-mass satellites are the only group for which the most important parameter for predicting quenching is environmental. This is in agreement with results obtained using global properties of simulated galaxies \citep{2024MNRAS.528.4891G,2025MNRAS.543.2006G} and results obtained from spatially resolved observations \citep{Bluck2020b}.

The bottom right panels of Figures \ref{fig:RF_classification} and \ref{fig:XGB_classification} show that $M_{\rm BH}$ is the parameter driving quenching in high-mass satellite galaxies, with $I_{\rm RF}(M_{\rm BH})=67\pm6\%$ and $I_{\rm XGB}(M_{\rm BH})=31\pm3\%$. Both Random Forest and \texttt{XGBoost} assign some importance to $\Sigma_*$ and $M_*$ ($I_{\rm RF}(\Sigma_*)=19\pm 5 \%$; $I_{\rm XGB}(\Sigma_*)=18\pm 2 \%$; $I_{\rm RF}(M_*)=13\pm 5 \%$; $I_{\rm XGB}(M_*)=18\pm 2 \%$). 
Both models find that quenching in high-mass satellites is driven by the same processes that drive quenching in central galaxies. However, the overall importance of black hole mass is less for the high-mass satellites than for the centrals, and there are larger variances between independent runs of the models for high-mass satellites. 

\end{sloppypar}


For the Random Forest models, training and testing AUCs range from $74 - 83\%$. For the \texttt{XGBoost} models, AUCs range from $78\%-83\%$. The differences between training and testing AUCs for most galaxy categories in both models is $\sim 3\%$,. We note that with a different combination of hyperparameters, better AUC scores could be obtained at the cost of an increased gap between training and testing, which is undesirable for the reasons described in \S \ref{sec:tuning}. 

In summary, we find that quenching in TNG100 is driven by global processes for all types of galaxies, in agreement with results from MaNGA observations \citep{Bluck2020b}.

\subsection{Regression}
\begin{figure*}[ht]
	\centering 
	\includegraphics[width=\linewidth]{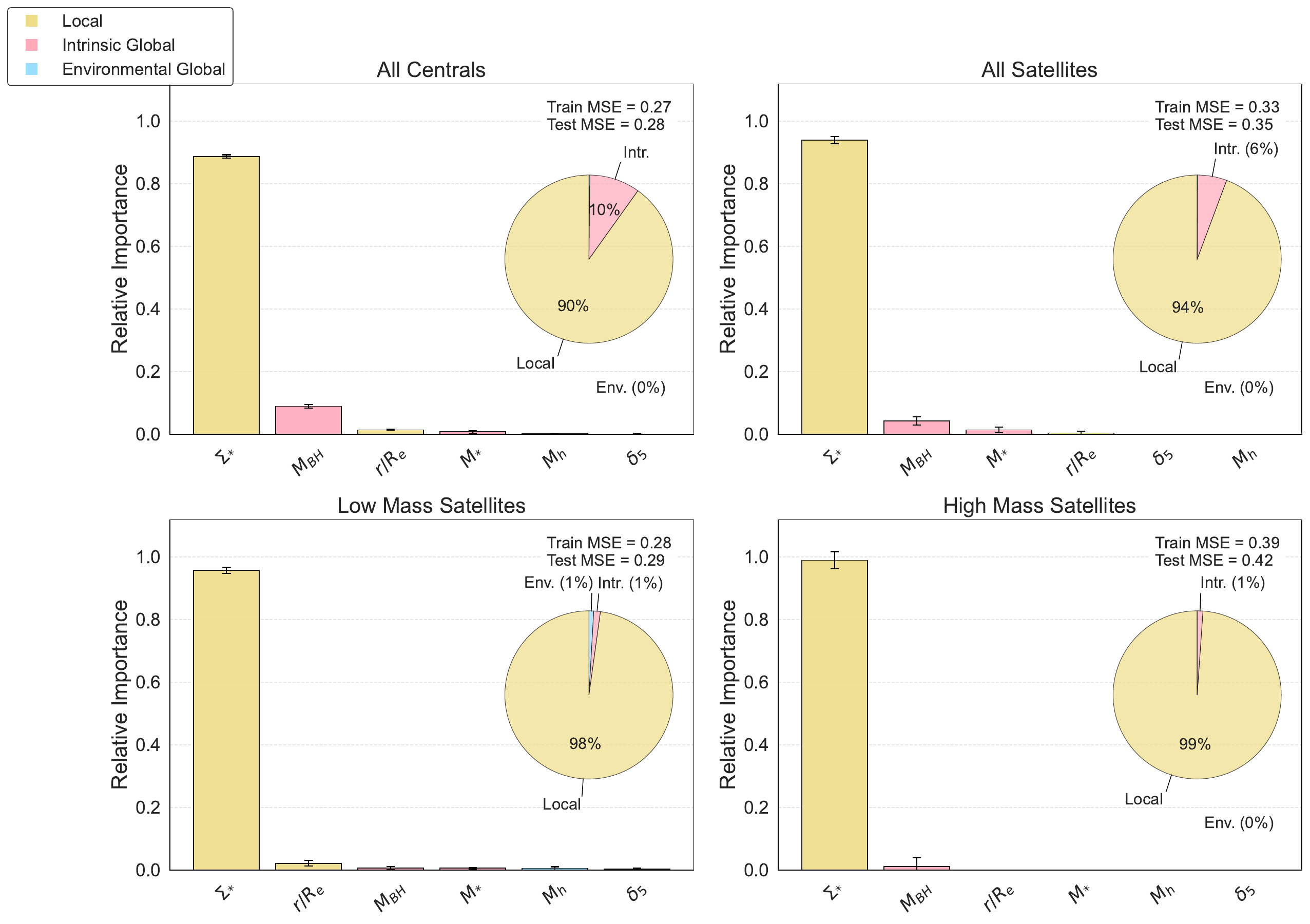}
	\caption{Feature importances from RF regression model for predicting SFR, shown for low-mass and high-mass subsets of both centrals and satellites. Formatting is the same as Figure \ref{fig:RF_classification}, except we report training and testing MSE scores for the regression models.} 
	\label{fig:RF_regression}%
\end{figure*}

\begin{figure*}[ht]
	\centering 
	\includegraphics[width=\linewidth]{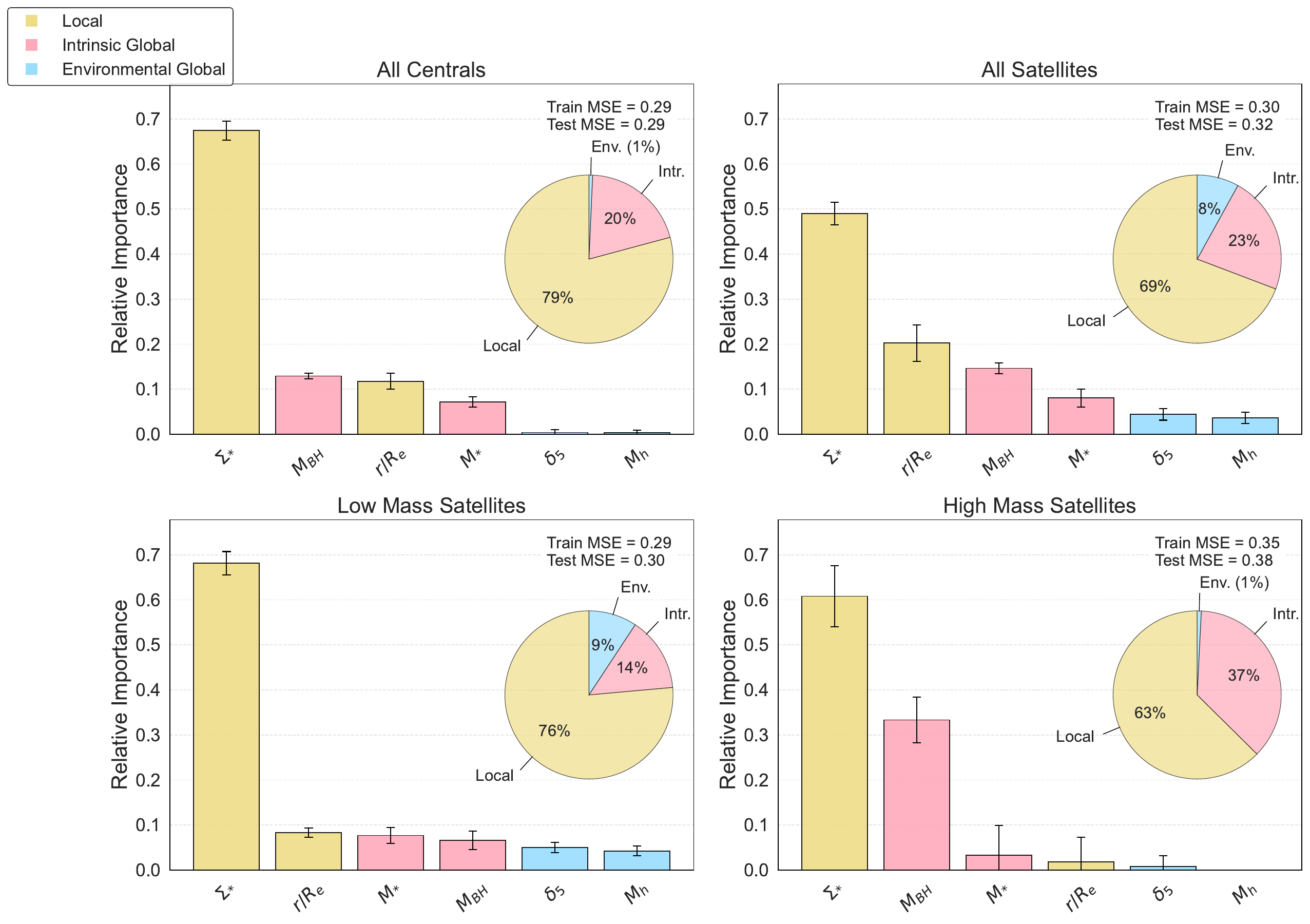}
	\caption{Feature importances from XGBoost regression model for predicting SFR, shown for low-mass and high-mass subsets of both centrals and satellites. Formatting is the same as Figure \ref{fig:RF_regression}} 
	\label{fig:XGB_regression}%
\end{figure*}


In Figures \ref{fig:RF_regression} and \ref{fig:XGB_regression}, we present the feature importances for predicting SFR obtained from Random Forest and \texttt{XGBoost} regression models, respectively. For these models, we use only the bins that are classified as star-forming and only those bins that come from galaxies classified as star-forming. 

For all galaxy subsets, the most important feature for predicting the $\Sigma_{\rm SFR}$ of a bin is its stellar mass surface density. In general, the relative importance of $\Sigma_*$ is higher in the Random Forest models than in the \texttt{XGBoost} models. It is not surprising that $\Sigma_{\rm SFR}$ is best predicted by $\Sigma_*$, since this is just the rSFMS \citep[e.g.,][]{Sanchez13,Cano-Diaz16,Liu_2018,Medling18,Erroz-Ferrer,Bluck2020a,McDonough2023ApJ...958...19M}. In \cite{McDonough2023ApJ...958...19M}, we showed that slope of the rSFMS recovered from TNG100 galaxies is in agreement with the slope recovered from MaNGA galaxies by \cite{Bluck2020a}. The random forest regression analysis performed on MaNGA data by \cite{Bluck2020b} found the same result we have here: $\Sigma_*$ is the best predictor of $\Sigma_{\rm SFR}$ of the analyzed variables. However, we note that our analysis is limited to the features analyzed. There could be another parameter, such as gas mass surface density, that would have a higher importance score than $\Sigma_*$ were it included.

For all central galaxies (Figures \ref{fig:RF_regression} and \ref{fig:XGB_regression}, top left), the relative importances of $\Sigma_*$ are $I_{\rm RF}(\Sigma_*)=88.7\pm 0.5\%$ and $I_{\rm XGB}(\Sigma_*)=67\pm 2\%$. In the Random Forest models, the only other feature with significant importance score ($I>5\%$) for central galaxies was black hole mass ($I_{\rm RF}(M_{\rm BH})=9.0\pm 0.6\%$). Black hole mass has a similar importance in the \texttt{XGBoost} models ($I_{\rm XGB}(M_{\rm BH})=12.9\pm 0.7\%$). However, $\Sigma_*$ and $r/R_e$ have similar importance scores ($\sim 10\%$) in the \texttt{XGBoost} models, albeit with larger variances.

We refrain from discussing in detail the results from all satellites, as there are significant differences between high- and low-mass satellites. For the low-mass satellites, $\Sigma_*$ is clearly the dominant feature ($I_{\rm RF}(\Sigma_*)=95.7\pm 0.9\%$; $I_{\rm XGB}(\Sigma_*)=68\pm 3\%$). For the \texttt{XGBoost} models, the remaining importance is divided relatively equally among the other features.

In the Random Forest models, the results for high-mass satellites are similar to those of all centrals ($I_{\rm RF}(\Sigma_*)=99\pm 3\%$), with even less importance assigned to $M_{\rm BH}$. However, this is reversed in the \texttt{XGBoost} models, where $I_{\rm XGB}(\Sigma_*)=61\pm7\%$ and $I_{\rm XGB}(M_{\rm BH})=33\pm5\%$

For the Random Forest models, training and testing MSEs range from $0.27 - 0.42 \; {\rm dex}^2$. For the \texttt{XGBoost} models, MSEs range from $0.29-0.38 \; {\rm dex}^2$. For both algorithms, the higher end of the range comes from the high-mass satellite population, which proved the hardest to train the models on. However, in all cases the testing MSE is smaller than the variance of the target variable.  The differences between training and testing MSEs for most galaxy categories in both models is $\sim 0.02$, with larger differences for high-mass satellites.

\begin{table*}[ht!]
\centering
\begin{tabular}{|p{2.5cm}|p{1.5cm}|p{1.5cm}|p{1.5cm}|p{1.5cm}|p{1.5cm}|}
\hline
 & & \multicolumn{2}{c|}{\textbf{Random Forest}} & \multicolumn{2}{c|}{\textbf{\texttt{XGBoost}}} \\
\cline{3-6}
\centering\textbf{Sample} & \centering var$(\Sigma_{\rm SFR})$ & \centering Test MSE & \centering $R^2$ & \centering Test MSE & \centering\arraybackslash $R^2$ \\
\hline
\textbf{All Centrals} & \centering $0.60$ & $0.28$ & $0.53$ & $0.29$ & $0.52$ \\
\hline
\textbf{All Satellites} & \centering $0.58$ & $0.35$ & $0.40$ & $0.32$ & $0.45$ \\
\hline
\textbf{Low-Mass Satellites} & \centering $0.59$ & $0.29$ & $0.51$ & $0.30$ & $0.49$ \\
\hline
\textbf{High-Mass Satellites} & \centering $0.56$ & $0.42$ & $0.25$ & $0.38$ & $0.32$ \\
\hline
\end{tabular}
\caption{Variance of $\Sigma_{\rm SFR}$ (dex$^2$), test MSE (dex$^2$), and $R^2$ for Random Forest and \texttt{XGBoost} regression models across all galaxy subsamples. }
\label{tab:regression_performance}
\end{table*}

To quantitatively assess how well the regression models have performed, we can compare the test MSE scores to the variance of the target variable, in this case $\Sigma_{\rm SFR}$. We compute the coefficient of determination for each population as:
\begin{equation}
    R^2=1-\frac{{\rm MSE_{test}}}{{\rm var (\Sigma_{SFR})}}.
\end{equation}
In Table \ref{tab:regression_performance}, we report the $R^2$ values for the regression models. A score of $R^2=0$ would indicate that the models are doing no better than the baseline prediction using the mean of $\Sigma_{\rm SFR}$, while $R^2=1$ would indicate that the models are predicting the correct value of $\Sigma_{\rm SFR}$ for every data point. For central and low-mass satellite galaxies, $R^2 \sim 0.5$, which indicates that the models for those systems can account for about $50\%$ of the natural variance of $\Sigma_{\rm SFR}$. We note again that the models for the high-mass satellite population perform much worse than the models for the other populations.


In summary, we find that star formation in TNG100 galaxies is regulated by local processes, while quenching is regulated by global process. This is identical to the conclusion drawn from a similar analysis of MaNGA galaxies \citep{Bluck2020b}.






\section{Discussion}
\label{sec:discussion}


In this work, we have applied the Random Forest and \texttt{XGBoost} approaches to classification and regression in order to determine the physical parameters that drive star formation and quenching within spatially-resolved regions of galaxies. The Random Forest approach has been used for this sort of investigation previously, and we will compare our results to these. 
\cite{Bluck2020a} and \cite{Bluck2020b} performed Random Forest classification and regression analysis on spatially resolved galaxies from MaNGA. The analysis of \cite{Bluck2020b} included separate models for all central galaxies, low-mass satellites, and high-mass satellites.
\cite{Piotrowska2022MNRAS.512.1052P} and \cite{2024MNRAS.528.4891G} both perform Random Forest classification analyses on galaxy-level data from observations (SDSS) and three simulations: TNG100, Illustris, and EAGLE. \cite{Piotrowska2022MNRAS.512.1052P} focused on central galaxies while \cite{2024MNRAS.528.4891G} analyzed low-mass and high-mass satellite galaxies. 
In the RF approach, we follow \cite{Bluck2020a} and \cite{Bluck2020b} in using Random Forest classification and regression to obtain relative feature importance scores. We refer readers to Appendix B of \cite{Piotrowska2022MNRAS.512.1052P} for tests of the robustness of these sorts of Random Forest analyses. We compare our results to a similar model that employs gradient-boosting, \texttt{XGBoost}. \texttt{XGBoost} is another popular machine learning tool for classification and regression, although to our knowledge it has not been used for the type of analysis we present here.
Both machine learning techniques offer a powerful new way to understand galaxy quenching by evaluating the relative importance of highly-correlated features for predicting quenching and SFR.

\subsection{Comparison of Random Forest and XGBoost results}\
\label{sec:comp_algos}

Results from both algorithms agreed on the most important feature for each category of galaxy explored. However, they often differ significantly in the magnitudes of the importance assigned to each feature. In general, compared to Random Forest, \texttt{XGBoost} assigns lower importance scores to the dominant feature and higher importance scores to other features. This phenomenon stems from fundamental differences in the underlying algorithms. 

The Random Forest algorithm builds decision trees independently and in parallel, with each tree evaluating all available features at each split. When features are highly inter-correlated (as is often the case for galaxies) and all features are evaluated at each split, Random Forest consistently selects the single strongest predictor across its ensemble of trees, resulting in concentrated importance scores. In contrast, \texttt{XGBoost} builds trees sequentially, with each subsequent tree attempting to correct the residual errors from previous trees through gradient-based optimization. After early trees exploit the dominant feature, the remaining prediction errors may be better explained by correlated features that contain overlapping but complementary information. This sequential error-correction process naturally distributes importance more evenly across correlated predictors, even when all features are available to each tree.

To interpret our results in light of these algorithmic differences between Random Forest and \texttt{XGBoost}, we argue that results from Random Forest are more reliable for determining the primary drivers of star formation and quenching given highly correlated parameters. That being said, \texttt{XGBoost} offers a complementary analysis that has potential for identifying secondary physical processes that contribute to quenching and star formation after accounting for the dominant driver. However, in our analysis the feature importance scores from independent runs of \texttt{XGBoost} typically have larger variances than those from RF runs. Because \texttt{XGBoost} trees are built sequentially in a given independent run, small variations in early trees (due to random subsampling) can lead the algorithm down different paths, resulting in each independent run identifying significantly different importance scores for secondary features. Critically, although we have ensured that \texttt{XGBoost} has all features \textit{available} at each split, variations in the early trees can result in the algorithm focusing on different features during different independent runs. This may result in large variance for redundant features (i.e., features with roughly equivalent predictive power). In contrast, independent Random Forest runs should converge on similar importance scores for redundant parameters, due to using independent decision trees. We have tested using more independent runs of the \texttt{XGBoost} models, but doing so does not meaningfully affect the resulting variance. 
For both Random Forest and \texttt{XGBoost}, large variance between independent runs can also result from poor sample size, true variation in evolutionary pathways for individual galaxies in a sample, or `missing features.' By missing features, we mean that there could be additional parameters that are not included in our analysis that would be better predictors of quenching. Such missing features would result in the models artificially inflating the importance of features correlated to the missing feature. 

In our conclusions below, we note that our analysis is limited by the features provided to the models. For example, we would not find that black hole mass is the dominant feature for predicting quenching of central galaxies if that feature were not included and other correlated features may be assigned higher importance scores. Likewise, the reported importance scores are \textit{relative} importance scores, because they will depend on the number and type of other features included. For this reason, we refrain from making quantitative comparisons of importance scores from other works. Instead, we focus on comparison of the features deemed most important.

\subsection{Quenching in central galaxies}
We find that black hole mass is the strongest predictor of quenching in bins from central galaxies ($I_{\rm RF}(M_{\rm BH})=75\pm2\%$; $I_{\rm XGB}(M_{\rm BH})=51\pm2\%$). This is in broad agreement with past works that have used RF \citep{Bluck2020a,Bluck2020b,Piotrowska2022MNRAS.512.1052P,2022A&A...659A.160B,2023ApJ...944..108B,2024ApJ...961..163B,2024MNRAS.528.4891G}. The mass of a galaxy's central supermassive black hole will be proportional to the amount of mass it has accreted, which in turn is proportional to the amount of energy released as AGN feedback \citep[see the appendix of][]{Bluck2020a}. Based on analysis of radial gradients in SFR and similar parameters, the current theory is that AGN feedback acts to quench a galaxy from the inside-out. Thus, it is unsurprising that we find black hole mass to be the dominant feature for predicting quenching. 

Our results for central galaxies are in agreement with the results from a similar Random Forest analysis of observed galaxies in MaNGA. Both \cite{Bluck2020b} and \cite{2022A&A...659A.160B} find that, when included, central velocity dispersion, the observational proxy for black hole mass, dominates the prediction of quenching in central galaxies. Thus, the TNG galaxy formation model appears to be sufficiently reproducing quenching of central galaxies as seen in observations.

In Random Forest classification analyses of central galaxies from simulations (EAGLE, TNG, Illustris) and observations (SDSS), \cite{Piotrowska2022MNRAS.512.1052P} and \cite{2024MNRAS.528.4891G} predicted quenching from galaxy-level data. We find similar results: $M_{BH}$ is by far the best predictor of quenching for central galaxies. In \cite{2023ApJ...944..108B,2024ApJ...961..163B}, this is found to be true up to $z\sim 4$.

By investigating the dependence of spatially resolved quenching on both global and spatially resolved parameters, we have expanded upon extant results to show that, as in observations, quenching is a global process in the TNG100 simulation.

\subsection{Quenching in low-mass satellites}
There are clear differences in the results for low-mass and high-mass satellite galaxies, so we will discuss them separately.

Several mechanisms have been proposed to explain the abundance of quenched, low-mass satellite galaxies in clusters. One is ram pressure stripping, which results in stripping of gas from an infalling satellite due to the large relative speeds between the satellite and intracluster medium. Satellites can also lose gas through tidal stripping or dynamical stripping, processes that will be correlated with local galaxy density \citep{1962AJ.....67..471K,1972ApJ...176....1G,2007A&A...472....5J,2009A&A...500..693J,2022A&ARv..30....3B,Wilkinson21}. 
These stripping mechanisms act to remove reservoirs of gas that could otherwise fuel star formation. In our analysis, we have examined the relative importance of halo mass and galaxy overdensity (measured at the $5^{\rm th}$ nearest neighbor) as proxies for ram pressure stripping and dynamical stripping/tidal stripping, respectively. 

We find that host halo mass is the best predictor of quenching in low-mass satellites, in both Random Forest and \texttt{XGBoost} ($I_{\rm RF}(M_h)=92\pm3\%$; $I_{\rm XGB}(M_h)=63\pm6\%$). In the Random Forest models, all other features have relative importance scores $\sim5\%$. Halo mass still dominates in the \texttt{XGBoost} models, but $M_{BH}$ has an importance score of $I_{\rm XGB}(M_{\rm BH})=34\pm4\%$. Thus, our results indicate that ram pressure stripping is the dominant quenching mechanism for low-mass satellite galaxies in TNG100. The \texttt{XGBoost} results may indicate that $M_{\rm BH}$ plays a secondary role in quenching low-mass satellites. 

Environmental parameters best predict the quenching of low-mass satellite galaxies in MaNGA data \citep{Bluck2020b}, and \cite{2024MNRAS.528.4891G} finds this to be the case with global data from SDSS and simulations. Our results are in agreement with \cite{2024MNRAS.528.4891G} in that we find halo mass to be the most important feature for predicting quenching in simulated low-mass satellites. However, both \cite{Bluck2020b} and \cite{2024MNRAS.528.4891G} find that local galaxy overdensity has a greater relative importance than $M_H$ for galaxies in observations. 

While \cite{2024MNRAS.528.4891G} performed their analysis on only integrated data from the simulation, we have extended this analysis to spatially resolved simulated data. As such, we are able to determine that quenching of low-mass satellites is not regulated locally but is instead a global process. That both our results and those of \cite{2024MNRAS.528.4891G} find a tension regarding the dominant quenching mechanism for low-mass satellites between contemporary simulations and observations indicates one of two likely possibilities: [1] TNG100 and the other simulations analyzed by \cite{2024MNRAS.528.4891G} (Illustris and EAGLE) fail to accurately model environmental quenching in low-mass satellites or [2] observed halo masses obtained from abundance matching techniques are not as correlated with the underlying quenching mechanism as observed local galaxy overdensity. Unlike AGN feedback, there are no subgrid prescriptions for environmental quenching; it is modeled directly by (magneto)hydrodynamics, which makes the first scenario unlikely. Although it is possible that there are important sub-resolution processes (e.g., instabilities that drive turbulence) that are not properly captured, this is unlikely given the scale that environmental effects operate on. This leaves the second explanation as the most likely, as halo masses derived from abundance matching techniques are poorly constrained.
Indeed, this is the conclusion drawn by \cite{2024MNRAS.528.4891G}, whose analysis indicates that local galaxy overdensity as measured at the $10^{\rm th}$ nearest neighbor may be a better predictor of halo mass than abundance matching techniques. 


This tension is a promising area for future studies, as it could lead to a better understanding of environmental quenching mechanisms and better estimates of observed halo masses, such as the method proposed by \cite{Bluck2025}. 

\subsection{Quenching in high-mass satellites}
\begin{sloppypar}
Compared to local and environmental features, intrinsic features better predict the quenching of high-mass satellites in both models ($I_{\rm RF}({\rm intrinsic})=80\pm11\%$; $I_{\rm XGB}({\rm intrinsic})=49\pm5\%$). In the Random Forest models, the dominate feature is $M_{\rm BH}$ ($I_{\rm RF}({\rm M_{\rm BH}})=67\pm6\%$), followed by $\Sigma_*$ and $M_*$ ($I_{\rm RF}({\rm \Sigma_*})=19\pm5\%$; $I_{\rm RF}({\rm M_*})=13\pm5\%$). In \texttt{XGBoost}, the dominance of $M_{\rm BH}$ ($I_{\rm XGB}({\rm M_{\rm BH}})=31\pm3\%$) is less clear, with $M_*$ and $\Sigma_*$ having importances of $\approx 18\%$, with $r/R_e$ and $\delta_5$ not far behind ($I_{\rm XGB}\approx 13\%$).

\end{sloppypar}

We note that the high-mass satellites proved to be the most challenging group when it came to hyperparameter tuning, showing consistent large gaps between training and testing AUCs. This indicates that the models consistently struggled to identify the feature that best predicted quenching. Thus, our results for this population are not as decisive as those for the others. There are several possibilities for why this is: 
there are true variations in evolutionary pathways for high-mass satellites, we are `missing feature(s),' or our sample of high-mass satellites is contaminated by intermediate-mass galaxies that do not quench in the same manner as their high-mass counterparts. Regarding the first point, there are results in the literature that point to stochasticity/anisotropy in the quenching of satellites \citep[e.g.,][]{2021Natur.594..187M,Simons20}. This stochasticity could result from variations in the orbital paths, AGN activity, and interaction history of satellites. In terms of `missing feature(s),' it is possible that there exists a property of high-mass satellites that explains the apparent stochasticity of quenching in these systems but which has not been explored in our analysis. In particular, in this analysis we have not here included any parameters related to galaxy morphology. Regarding the final point, 
\cite{2024MNRAS.528.4891G} did not find much ambiguity in the primary driver of quenching in high-mass satellites when they defined high-mass satellites as $M_*>10^{10.5} M_\odot$. They found that $M_{\rm BH}$ was clearly the most predictive feature for quenching in high-mass satellites from TNG and SDSS. \cite{Bluck2020b} also found this to be the case (using central velocity dispersion as a proxy for $M_{\rm BH}$) for spatially resolved data from MaNGA satellites with $M_*>10^{10.5} M_\odot$. Our adoption of a mass cut at $2\times 10^{10} M_\odot$ was motivated by the quenching mass threshold identified by \cite{2025MNRAS.543.2006G}. However, it is possible that quenching of satellites near this mass threshold is more complicated than for satellites with masses much less than or much greater than the threshold. 

If it were the case that our results for high-mass satellites were being influenced by the presence of an intermediate-mass population, that population does not appear to quench in a similar manner to low-mass satellites because $M_h$ is the \textit{least} important feature for high-mass satellites. Compare this to the top right panel of Figure \ref{fig:RF_classification}, where the mixed `all satellites' population shows a nearly equal split between $M_h$ and $M_{\rm BH}$. However, both algorithms detect $M_*$ as an important feature, so we cannot rule out the possibility that there is contamination by intermediate-mass satellites in the sample.

In theory, high-mass satellites share properties with both central galaxies (which tend to be more massive) and low-mass satellite galaxies (in that they are both orbiting a gravitational potential well). So it may be expected that high-mass satellites can quench through both AGN feedback and environmental processes. Indeed, radial profiles of SFR and similar properties for high-mass satellites can show both inside-out and outside-in (i.e., environmental) quenching \citep{Bluck2020b,McDonough2025ApJ...978...67M}. However, our results do not indicate that any environmental features are particularly predictive of quenching for high-mass satellites. Indeed, high-mass galaxies are better able to resist gas stripping due to their larger gravitational potentials. We leave the exploration of the ambiguity regarding the quenching of high-mass satellite galaxies in TNG100 to future work.

\subsection{Drivers of star formation}
While the quenching of star formation appears to be a global processes driven by intrinsic or environmental processes, we find that star formation itself is best predicted by local features. In particular, both Random Forest and \texttt{XGBoost} regression assigns the most importance to stellar mass surface density ($\Sigma_*$) for all categories of galaxies. This is as expected given the relationship between $\Sigma_{\rm SFR}$ and $\Sigma_*$ described by the rSFMS \citep[e.g.,][]{Sanchez13,Cano-Diaz16,Liu_2018,Medling18,Erroz-Ferrer,Bluck2020a,McDonough2023ApJ...958...19M}. For all populations, Random Forest assigns $\Sigma_*$ importance scores of $\sim 90\%$.

For the \texttt{XGBoost} models, $\Sigma_*$ is still very clearly dominate, with $I_{\rm XGB}(\Sigma_*)\gtrsim60\%$ for all centrals, low-mass satellites, and high-mass satellites. Most other features have importance scores $\lesssim20\%$, with the exception of $M_{\rm BH}$ for high-mass satellites, with $I_{\rm XGB}(M_{\rm BH})=33\pm 5\%$.


Our results are in general agreement with those of \cite{Bluck2020a}: star formation is driven by local processes, while quenching is driven by global processes. This result indicates that the TNG galaxy formation model is reproducing, on both global and spatially resolved scales, the processes that regulate star formation as identified in observations. Here and in observations, $\Sigma_*$ serves as a proxy for the presence of molecular gas from which stars may form. The importance assigned to $M_{\rm BH}$ for some populations indicates that $\Sigma_*$ is an imperfect proxy, and this feature provides the models with the information they need to differentiate between bins with similar stellar masses but differing SFRs. For instance, large $M_{\rm BH}$ will be associated with greater AGN feedback, which will act to reduce SFR by expelling gas or decreasing star formation efficiency. 


\subsection{Principal Component Analysis}
In order to further support our results, we conduct a more traditional principal component analysis (PCA) that does not rely on machine learning. We focus specifically on our finding that quenching is driven by global processes while star formation is regulated by local processes. We reduce the dimensionality of this problem by independently computing the first principal components of the local parameters ($\Sigma_*$ and $r/R_e$) and the global parameters ($M_{\rm BH}$, $M_*$, $M_h$ and $\delta_5$). We compute these hyper-parameters, PC$_1$(local) and PC$_1$(global), from all annular bins, regardless of classification as central or satellite, low-mass or high-mass.

To compute the local and global hyper-parameters, we first scale the data by taking the base-10 logarithm. Because $\delta_5$ can have negative values, we add one to this parameter before taking the logarithm. For $M_{\rm BH}$, we replace any zero values (where a black hole was not seeded), with half of the seed mass. For each parameter, we then subtract the mean and normalize by the standard deviation. This has the effect of centering the data and ensuring each parameter follows a similar scale. We adopt the \texttt{scikit-learn.decomposition.PCA} class to perform the standard eigenvector/eigenvalue decomposition in order to determine a projection of the multidimensional data which is aligned with directions of maximal variance. Our hyper-parameters are the first principal component of these decompositions performed on the global and local data separately. These hyper-parameters effectively reduce our multi-dimensional problem to a two-dimensional problem, where we can directly compare the influence of local variables versus global variables. We note that the PCA eigenvectors have no intrinsic sign, so we flip the sign of PC$_1$(local) such that it increases with increasing $\Sigma_{*}$, consistent with \cite{Bluck2020a}.

\begin{figure*}
    \centering
    \includegraphics[width=\linewidth]{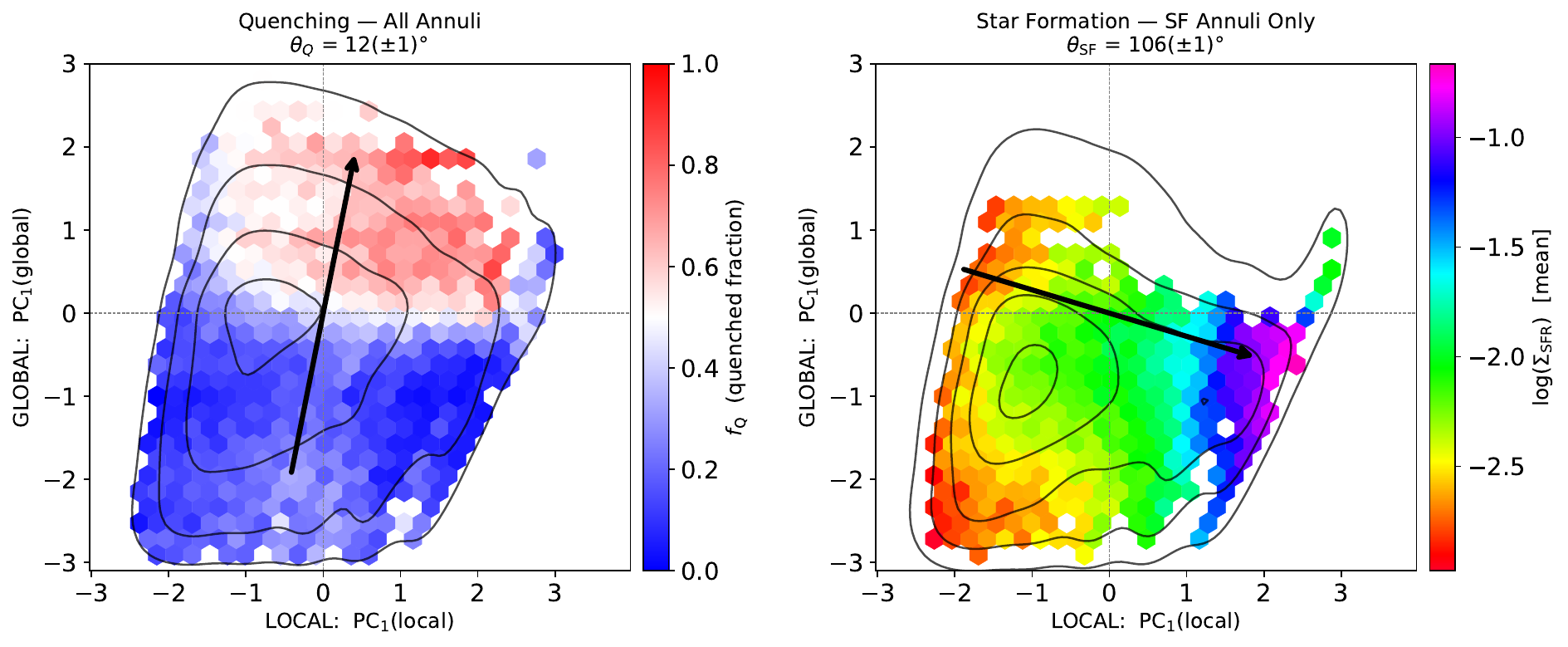}
    \caption{Principal component analysis of local and global parameters. In both panels, we plot the global hyperparameter, PC$_1$(global) versus the local hyperparameter, PC$_1$(local). Contours trace the number density of annular regions. At left, the hexagons are colored based on the fraction of quenched annular regions contained in each bin ($f_Q$). At right, the hexagons are colored based on the mean star formation rate density ($\Sigma_{\rm SFR}$) of annular regions contained in each bin. Only star-forming annular regions are considered in the right panel. We also draw the axial ratio statistic ($\theta_Q$ or $\theta_{\rm SF}$), which shows the direction of steepest increase in $f_Q$ or $\Sigma_{\rm SFR}$ for the left and right panels, respectively.  }
    \label{fig:PCA}
\end{figure*}

In Figure \ref{fig:PCA}, we plot the relationship between PC$_1$(global) and PC$_1$(local) for all annular regions (left) and star-forming annular regions (right). The density contours indicate how annular regions are distributed on this plane. In the left-panel we color-code the plot by the fraction of quenched annular regions ($f_Q$) contained within each hexagonal bin. It is clear that there is little trend between the local hyperparameter and $f_Q$, but that $f_Q$ increases with increasing PC$_1$(global). This indicates that, for the TNG100 galaxies we examine, quenching is driven by global, not local, processes. 

Likewise, Figure \ref{fig:PCA} (right), shows the PC$_1$(global)-PC$_1$(local) plane for star-forming regions, with hexagonal bins color-coded by the mean star formation rate surface density ($\Sigma_{\rm SFR}$). Here, it is clear that $\Sigma_{\rm SFR}$ varies little with varying global parameters, but is strongly correlated with local parameters. Again, this result is expected given the resolved star formation main sequence.

\cite{Bluck2020a} introduces the axial ratio statistic to quantify these observations. We briefly summarize the methodology here, but refer readers to \cite{Bluck2020a} for further details. We compute the partial correlation coefficient (PCC) to quantify the strength of correlation between two variables (e.g., one of the hyper-parameters and the color bar parameter) at a fixed third variable (the other hyper-parameter). If (for example) $\rho_{AB}$ is the Spearman rank correlation coefficient between variables $A$ and $B$, then the PCC is defined as:
\begin{equation}
    \rho_{AB,C}= \frac{\rho_{AB}-\rho_{AC} \rho_{BC}}{\sqrt{1-\rho_{AC}^2} \sqrt{1-\rho_{BC}^2}}.
\end{equation}

Defining $X={\rm PC}_1({\rm local})$, $Y={\rm PC}_1({\rm global})$, and $Z$ as the color bar variable ($f_Q$ or $\Sigma_{\rm SFR}$), we compute two PCCs for each panel of Figure \ref{fig:PCA}: $\rho_{XZ,Y}$ and $\rho_{YZ,X}$. Treating the two PCCs as components of a vector, we can then define an angle (clockwise from vertical) which represents the direction through the plane which maximizes $f_Q$ or $\Sigma_{\rm SFR}$. We thus adopt the axial ratio statistics:
\begin{equation}
    \theta_Q=\tan^{-1}(\frac{\rho_{Y f_Q, X}}{\rho_{X f_{Q}, Y}})
\end{equation}
and
\begin{equation}
    \theta_{SF}=\tan^{-1}(\frac{\rho_{Y \log\Sigma_{\rm SFR}, X}}{\rho_{X \log\Sigma_{\rm SFR}, Y}}).
\end{equation}

A value of $90 \degree$ would indicate that the quenched fraction (or $\Sigma_{\rm SFR}$) is entirely correlated with PC$_1$(local) and entirely uncorrelated with PC$_1$(global). Likwise, a value of $0 \degree$ for the axial ratio statistic would indicate that the quenched fraction (or $\Sigma_{\rm SFR}$) is entirely correlated with PC$_1$(global) and entirely uncorrelated with PC$_1$(local). 

We estimate the uncertainties on $\theta_Q$ and $\theta_{SF}$ by taking the standard deviation of the results from $1000$ bootstrap resamplings of galaxies (not annular bins). In each resampling, the PCCs and axial ratio statistic are re-computed.

We find $\theta_Q=12\pm1\degree$ and $\theta_{SF}=106\pm1 \degree$. These results are in agreement with our qualitative assessment as well as our results from the machine learning analysis. That is, quenching is mostly driven by global processes while star formation is mostly driven by local processes. 

If quenching were a natural conclusion of the processes driving star formation, we might expect that $\theta_{SF}-\theta_Q=180\degree$. (That is, that the vector maximizing $\Sigma_{\rm SFR}$ should be opposite the vector minimizing $f_Q$. However, with $\theta_{SF}-\theta_Q=94\pm2\degree$, this is clearly not the case. In fact, the two vectors are nearly perpendicular, so we may conclude that quenching and star formation are driven by different processes.

\cite{Bluck2020a} performed this analysis for central galaxies observed by MaNGA, although their global and local hyper-parameters were constructed from observed properties, so a direct quantitative comparison is not reasonable. They find $\theta_Q=16\pm4\degree$ and $\theta_{SF}=93\pm3\degree$. Despite significant differences in the data underlying our analyses, we can thus conclude that the processes that drive quenching and star formation in TNG100 galaxies are similar to the processes operating on observed galaxies. That is, in both observations and simulations, star formation is governed by local processes and quenching is governed by global processes.

\section{Summary and conclusions}
\label{sec:summary}

We have trained Random Forest and \texttt{XGBoost} classification models to predict whether regions of galaxies will be quenched or star-forming based on features associated with that region and the region's host galaxy. Likewise, we have also trained Random Forest and \texttt{XGBoost} regression models to predict $\Sigma_{\rm SFR}$ for individual bins. These algorithms assign importance scores to the provided features based on how important a given feature is for making the predictions. We interpret the relative importance of features to be indicative of processes that drive star formation or quenching for the simulated TNG100 galaxies in our sample. 

We emphasize that our sample consists of simulated galaxies, which means that the conclusions summarized below are sensitive to the limitations of simulated data (e.g., resolution, sub-grid models, etc.). However, our data set of spaxels from TNG100 galaxies has been shown to reproduce the resolved star-formation main sequence and the expected radial distribution of $\Sigma_*$ and luminosity-weighted age \citep{McDonough2023ApJ...958...19M}.
The goal of this work is to establish the mechanisms underlying quenching and regulation of star formation within simulated galaxies to compare against observations. Our main conclusion is thus: for any galaxy, regardless of mass or status as central or satellite, quenching is driven by global processes while star formation is driven by local processes. Our other key results are as follows:

\begin{itemize}
    \item Random Forest (RF) and \texttt{XGBoost} models generally agree in their ranking of feature importance (within variance). Quantitatively, however, \texttt{XGBoost} tends to assign a lower importance than RF to the most important feature and distributes that importance over less important features. This is a result of the different algorithms these methods use. For the purpose of understanding galaxy evolution, RF offers a simple way to establish the most important parameter among highly-correlated features. Our application of RF to this problem has benefited from lessons learned from similar analyses \citep{Bluck2020a, Bluck2020b,Piotrowska2022MNRAS.512.1052P,2024MNRAS.528.4891G}, in particular the adoption of \texttt{max\_features=None}. \texttt{XGBoost} may be a complementary analysis tool, but more work is necessary to understand how the algorithm handles highly-correlated features, such as the analysis in Appendix B of \cite{Piotrowska2022MNRAS.512.1052P} or Appendix B of \cite{2022A&A...659A.160B}.

    \item For predicting the surface density of SFR in a given spaxel, $\Sigma_*$ is the most important feature regardless of the type of galaxy. Features unique to galaxies (rather than bins) are the most important for predicting whether a given bin is star forming or quenched. This indicates that star formation is a local process, while quenching is a global process. This is the same conclusion reached by \cite{Bluck2020b} using observational data and supported by our PCA results (Figure \ref{fig:PCA}). Hence, the TNG galaxy formation model is reproducing quenching and star formation regulation on kiloparsec-scales, as seen in observations.

    \item For central galaxies, we find that $M_{\rm BH}$ is by far the most important predictor of whether a bin will be quenched or star-forming. This is consistent with the results found in observations \citep[e.g.,][]{2016MNRAS.462.2559B,Piotrowska2022MNRAS.512.1052P,2022A&A...659A.160B}. 

    \item High-mass satellites are similar to central galaxies in 
    that the most important feature for predicting quenching is $M_{\rm BH}$. However, the importance score for this feature is lower in high-mass satellites, especially in the \texttt{XGBoost} models. This may indicate that, while $M_{\rm BH}$ is the main factor driving quenching in high-mass satellites, other processes may be operating.

    \item The results for low-mass satellites are unambiguous: host halo mass is the most important feature for predicting quenching of a given spaxel. The importance scores assigned to $M_h$ by \texttt{XGBoost} (RF) are a factor of $\sim10$ ($\sim100$) than those assigned to $\delta_5$. This indicates that quenching of low-mass satellites is driven by galaxy-halo interactions, rather than galaxy-galaxy interactions. This is consistent with other results derived from TNG100 galaxies \citep{McDonough2025ApJ...978...67M}, but in tension with a similar analysis of MaNGA galaxies by \cite{Bluck2020b}. This tension between results from observations and those from simulations was also found by \cite{2024MNRAS.528.4891G}. The source of this tension may lie in limitations to galaxy evolution simulations but is likely due to how halo mass is inferred from observations, as indicated by the results presented in \cite{2024MNRAS.528.4891G}. However, further investigation into this tension is desirable. 

\end{itemize}

In future works, we intend to explore within TNG100 galaxies the relative importance of gas availability versus star formation efficiency, investigate the importance of available morphological features in driving quenching, and further explore the tension regarding environmental quenching of low-mass satellite galaxies.

\section*{Declaration of generative AI and AI-assisted technologies in the manuscript preparation process}
During the preparation of this work the authors used Claude.ai to draft a few technically complex passages in order to ensure accuracy and clarity regarding the discussed algorithms. The abstract was also initially drafted by Claude.ai, which was provided with the rest of the manuscript. After using this tool/service, the authors reviewed and edited the content as needed and take full responsibility for the content of the published article.

\section*{Acknowledgments}
We thank the IllustrisTNG team for producing and providing the simulated data. The IllustrisTNG simulations were undertaken with compute time awarded by the Gauss Centre for Supercomputing (GCS) under GCS Large-Scale Projects GCS-ILLU and GCS-DWAR on the GCS share of the supercomputer Hazel Hen at the High Performance Computing Center Stuttgart (HLRS), as well as on the machines of the Max Planck Computing and Data Facility (MPCDF) in Garching, Germany. We also thank the teams behind the software we have used to complete this work, in particular the team behind the \texttt{scikit-learn} Python package. This work was completed in part using the Explorer Cluster, supported by Northeastern University’s Research Computing team.

\textit{Funding}: B.M. acknowledges support by Northeastern University's Future Faculty Postdoctoral Fellowship Program and thanks the ADVANCE office for the opportunity to pursue this work. A.F.L.B. gratefully acknowledges support from an NSF research grant: NSF-AST 2408009, in addition to an ORAU Junior Faculty Enhancement Award in Physical Sciences, and research start-up funds from FIU.

\section*{CRediT authorship contribution statement}
\textbf{Bryanne McDonough}: Conceptualization, validation, formal analysis, data curation, writing - original draft, writing - review \& editing, project administration. \textbf{Sathvika S. Iyengar}: Software, formal analysis, investigation, writing - original draft, writing - review \& editing, visualization. \textbf{Ansa Brew-Smith}: Software, formal analysis, investigation, writing - review \& editing, visualization. \textbf{Asa F. L. Bluck}: Conceptualization, methodology, writing - review \& editing, supervision. \textbf{Joanna M. Piotrowska}: Methodology, writing - review \& editing, supervision.




\bibliographystyle{elsarticle-harv} 
\bibliography{main}






\end{document}